\documentclass[epj]{svjour}
\usepackage{graphics}
\usepackage{color}
\usepackage{amsmath,amssymb}
\usepackage{hyperref}
\newcommand{\df}[2]{ \frac{\partial {#1}}{\partial {#2}} }

\begin{document}
%
%\title{Effect of strong magnetic fields on compact star structure}
\title{Structure of ultra-magnetised neutron stars}
\author{Debarati Chatterjee\inst{1} \thanks{\emph{debarati@iucaa.in}}, J\'er\^ome Novak\inst{2} \and Micaela Oertel\inst{2}% etc
}                     % Do not remove
\offprints{}          % Insert a name or remove this line
\institute{Inter-University Centre for Astronomy and Astrophysics, \\
Post Bag 4, Ganeshkhind, Pune University Campus,\\
Pune - 411007, India 
\and 
LUTH, Observatoire de Paris, \\
Universit\'e PSL, CNRS, Universit\'e de Paris,\\
92195 Meudon, France}
\date{Received: date / Revised version: date}
\abstract{
In this review we discuss self-consistent methods to calculate the global structure of strongly magnetised neutron stars within the general-relativistic framework. We outline why solutions in spherical symmetry cannot be applied to strongly magnetised compact stars, and elaborate on a consistent formalism to compute rotating magnetised neutron star models. We also discuss an application of the above full numerical solution for studying the influence of strong magnetic fields on the radius and crust thickness of magnetars. The above technique is also applied to construct a ``universal'' magnetic field profile inside the neutron star, that may be useful for studies in nuclear physics. The methodology developed here is particularly useful to interpret multi-messenger astrophysical data of strongly magnetised neutron stars. 
\PACS{
      {PACS-key}{describing text of that key}   \and
      {PACS-key}{describing text of that key}
     } % end of PACS codes
} %end of abstract
\maketitle
\section{Introduction}
\label{intro}
The phase diagram of Quantum Chromodynamics (QCD), that describes the
behaviour of matter at different densities and temperatures, has
intrigued physicists since de\-cades. While low density and low
temperature physics is probed in terrestrial nuclear experiments,
finite temperatures and densities are accessible to heavy-ion
collision experiments. High temperature and low density physics is of
relevance to the early Universe and can be probed with techniques such a
Lattice QCD. Neutron stars (NS), on the other hand, allow us to
investigate the high density low temperature regime of the phase
diagram that is complementary to terrestrial experiments. Hot
proto-neutron stars newly born from core collapse supernovae or binary
neutron star mergers lie in the finite temperature part of the phase
diagram. The understanding of the nature of dense matter encompasses
many different disciplines in physics. Comparison of information from
the different regimes of the phase diagram provides important input in
understanding the behaviour of dense matter under extreme
conditions. 

Another important question in this regard is the nature of the QCD
phase diagram under strong magnetic fields. An important insight can
come from observations of neutron stars (pulsars) endowed with strong
magnetic fields. In general, standard pulsars are known to have large
surface magnetic fields around $10^{10}-10^{12}$ G. This can be
explained by the conservation of flux of the progenitor star as it
collapses to a neutron star via a core-collapse supernova
explosion. Further, astrophysical observations indicate the existence
of a class of neutron stars possessing ultra-strong magnetic fields, commonly referred to as {\it magnetars} (see e.g. the pioneering articles~\cite{Duncan92,Usov92,Paczynski} and the review~\cite{Kaspi} for a comprehensive understanding). It is commonly thought that magnetars are observed as Anomalous X-ray Pulsars (AXPs) and Short Gamma-ray Repeaters (SGRs). From the characteristic common features such as eruptive emission of X-rays and gamma rays, it has been concluded that the energy associated with these bursts of intense radiation~\cite{CotiZelati} is probably due to ultra-strong magnetic fields, with surface fields reaching $10^{15}-10^{16}$ G. Recently, isolated neutron stars, such as X-ray Dim Isolated NSs (XDINSs) and Rotating Radio Transients (RRATs) with intermediate magnetic field strengths have also been observed \cite{Popov}. Multi-messenger observations of the compact remnant resulting from the neutron star merger event GW170817 also suggest that it may be a hypermassive differentially rotating magnetar \cite{Tong,Ai,Gill}. Similarly, the observation of GRB~200522A was associated with the formation of a stable magnetar, in order to explain some of the afterglow emission \cite{Fong}. Thus for the correct interpretation of astrophysical data from strongly magnetised neutron stars, it is crucial to develop accurate models of the structure of such objects.

 This article is organised as follows: in Sec.~\ref{sec:maxfield} we briefly recall the maximal value of the magnetic field inside compact stars, before discussing in Sec.~\ref{sec:effects} its effects on star's microscopic and macroscopic properties. Sec.~\ref{sec:model} describes the numerical approach to get numerical models of strongly magnetised compact stars, which is then applied in Sec.~\ref{sec:mageos} to compute the effect of a magnetised equation of state on their structure. In Sec.~\ref{sec:magprofile} we devise a so-called universal profile for magnetic field distribution in compact stars, before giving in Sec.~\ref{sec:conc} some concluding remarks. Unless otherwise stated (in particular when discussing magnetic field values) we use natural (Dirac) units such that $G=c=\hbar=1$.

%%% Origin of magnetars or High-B NSs??
%%magnetised WDs, Supra-Chandrasekhar mass WDs.??
%%Describe phenomena like magnetar glitches, QPOs?

\section{Maximum interior magnetic field}
\label{sec:maxfield}
Several observations can be used to estimate the surface magnetic
fields in magnetars. From the measurements of pulsar rotation periods
and period derivatives (so-called $P-\dot{P}$ diagram), assuming a
dipole model for the magnetic field, one can estimate the magnetic
fields to be $\sim 10^{15}-10^{16}$~G for some objects. Direct
estimates of the field on the surface of magnetars is also
possible through the observation of cyclotron lines
\cite{Ibrahim,Mereghetti}. 

However, there are no astrophysical observations to directly measure the magnetic field at the centre of a neutron star. Applying the Virial Theorem (comparison of magnetic energy with that of matter and gravitational potential energy)~\cite{ChandraFermi,Shapiro,Bonazzola94,Gourgoulhon94}, the estimate of the maximum magnetic field strength in the interior of a magnetar turns out to be  $\sim 10^{18}$~G. However, the exact estimate of the maximum central field in this way is also complicated, as these energy contributions in turn also depend on magnetic field.
%% White dwarfs ??

%%=============================================
\section{Effects of strong magnetic field}
\label{sec:effects}

Ultrastrong interior magnetic field strengths $\sim 10^{18}$~G may affect compact stars mainly in two ways: 
\begin{itemize}
    \item The magnetic field interacts with the particles in the stellar interior, thus affecting the Equation of State (EoS)
    \item Strong magnetic fields result in a modification of the energy-momentum tensor, breaking of the spherical symmetry, and therefore affect the external structure of the compact star.
\end{itemize}
We elaborate on these aspects in detail in the following sections.

\subsection{Effect on the Equation of state}
\label{sec:int}
Since for neutron stars older than several minutes the typical temperatures are negligible for the equation of state, we will consider here only the $T = 0$ case. In addition we assume matter to be in chemical ($\beta$) equilibrium.
In the NS interior, charge conservation and chemical equilibrium result in the appearance of a small fraction of protons and electrons, in addition to neutrons. Further, as the baryon densities in the NS interior surpass $\sim$ 2-3 times that of nuclear saturation density ($n_0 \sim 2.3 \times 10^{17}$ kg/m$^3$), strangeness containing particles such as hyperons, condensates of kaons or even deconfined quarks may appear there (see Ref.~\cite{Haensel2007} for detailed discussions). The appearance of these particles depends on an interplay between the neutron and electron chemical potentials that govern the weak interactions that produce them. The composition of the NS core affects the relation between the matter pressure and energy density (EoS), see e.g.~\cite{ChatterjeeVidana}.

In presence of a magnetic field, the motion of charged particles
becomes confined to quantised Landau levels \cite{Landau} and on one hand the EoS
is affected by the Landau quantisation of the charged
particles (such as protons, electrons, etc). The effects are most pronounced when the particle is confined to the lowest Landau level. Additionally, as first studied by Canuto and Chiu \cite{Canuto} in the case of a relativistic electron gas, the energy-momentum tensor becomes anisotropic. On the other hand, the
interaction of the magnetic moments, including the anomalous magnetic
moments of the neutral particles (such as neutrons,
$\Lambda$-hyperons, \dots), with the magnetic field influences the
EoS.

The effect of quantising magnetic fields on the EoS for different models of NS matter can be found in many articles in the literature, see e.g.~\cite{Bandyopadhyay,Chakrabarty,Broderick00,Broderick02,Strickland,Noronha,Rabhi,Ferrer,Sinha}.
In presence of a uniform external magnetic field in the $z$ direction, the transverse momenta of particles (electrons, protons, hyperons) are restricted to discrete Landau levels with squared transverse momenta $k_{\perp}^2 = 2 \nu q b$ where $ \nu \ge 0$ is the Landau quantum number. For spin-1/2 particles, $\nu$ is related to the orbital angular momentum $n$ by\footnote{For a derivation of the following equations, see e.g. the textbooks \cite{Haensel2007}, chapter 4, or \cite{Landau}.}~  
\begin{equation}
\nu = n +\frac{1}{2}  - \frac{s}{2} \frac{q}{\lvert{q}\rvert}~,
\end{equation}
where $s \pm 1$ is the spin projection of the particle in the direction of the magnetic field. The total energy of a charged particle becomes quantised as (see~\cite{Broderick00}):
\begin{equation}
E = \sqrt{k_z^2 + \left[(m^2 + 2\nu \lvert{q}\rvert b)^{1/2} - s \kappa b \right]^2} 
= \sqrt{k_z^2 + \bar{m}^2(\nu)}~,  
\label{eq:energy_mag}
\end{equation}
$k_z$ being the momenta in the $z$-direction and $\kappa$ the anomalous magnetic moment \cite{Strickland}, while 
\begin{equation}
  \bar{m}^2(\nu) = (\sqrt{m^2 + 2 \nu \lvert q \rvert b} - s \kappa b )^2 ~.
  \label{eq:mbar}
  \end{equation}
If one considers the simplest case with $\kappa=0$, the maximum momentum $k_z$ in terms of the chemical potential $\mu$ is defined as:
\begin{equation}
k_{z,F} = \sqrt{\mu^2 - 2 \nu \lvert q \rvert b - m^2}~.
\end{equation}
To ensure that the quantity under the square root is positive, one may impose the condition:
\begin{equation}
\nu \leq \nu_{max} = \left[ \frac{\mu^2 - m^2}{2 \lvert q \rvert b} \right]~.
\end{equation}
A magnetic field strong enough such that only the lowest Landau level
is occupied is in general called a strongly quantising field.  Using
the above relations, one can write down the expression for the number
density as
\begin{equation}
n = \frac{\lvert q \rvert b}{2 \pi^2} \sum_{s=\pm1} \sum_{\nu=0}^{\nu_{max}} k_{z,F}(\nu)~.
\end{equation}
Similarly, one can obtain the energy density as
\begin{equation}
\epsilon = \frac{\lvert q \rvert b}{2 \pi^2}  \sum_{s=\pm1} \sum_{\nu=0}^{\nu_{max}} \int_0^{k_{z,F}} d k_z \sqrt{k_z^2 + \bar{m}^2(\nu)}~.
\end{equation}
From the energy density, one may derive the pressure using the
Gibbs-Duhem relation, and therefore obtain the EoS. The situation is
more complicated in the presence of non-zero anomalous magnetic moment
\cite{Strickland}, and the full expression in Eq.~\eqref{eq:energy_mag} has to be used.

As mentioned above, these effects have been included in numerous
studies of the neutron star matter EoS in a strong magnetic field.  In
the neutron star crust, due to the magnetic field electronic motion is
quantised into Landau levels. For field strengths $b \gg
b^e_c$, the crust composition depends on the magnetic field strength
with typical quantum oscillations observed in the transition density
from one nuclear cluster to another. $b^e_c = m_e^2/|e| \sim 4.4 \times 10^{13}$~G denotes here the critical field for electrons~\cite{Haensel2007}. If the magnetic field strength
exceeds the value for being strongly quantising, this transition density is
increasing linearly with the magnetic field strength, leading to a
less neutron rich crust and an increased density for neutron drip, 
i.e. shifting the transition from the outer to the inner crust to
higher densities~\cite{Chamel2015}. In the transition region from the inner to the outer crust, field strengths above roughly $\sim 1300 b^e_c$ correspond to strongly quantising fields. In the crust, for magnetic fields above a few times
$10^{17}$ G, nuclear binding energies start to become significantly
altered, modifying additionally the structure of the crust, see
e.g.~\cite{Pena}. A summary of crust properties under strong magnetic
fields can be found in Ref.~\cite{BlaschkeChamel}.

The crust-core interface is characterised by the transition from inhomogeneous matter in the crust to homogeneous matter in the core. Ravenhall et al.~\cite{Ravenhall1984} and Hashimoto et al.~\cite{Hashimoto1984} predicted that this transition passes via more and more deformed nuclei, forming the so-called "nuclear pasta", with the detailed structure depending on the nuclear interaction. Meanwhile, 
different calculations and numerical techniques have been used to
investigate the crust-core interface region, see e.g.\cite{Oyamatsu,Pethick,Watanabe,Avancini2008,Ducoin2010,Avancini2010,Providencia2017}, confirming essentially the above picture and obtaining comparable
values for crust-core transition densities. The crust-core region of
magnetars was recently studied within the relativistic mean field
model framework \cite{Fang2016,Fang2017}. In these works, the effects
of strong magnetic fields ($10^{15}-10^{17}$ G) on the instability
region was investigated with the help of the Vlasov formalism used to
determine the dynamical spinodals. It is found that again, since the
inner crust becomes less neutron rich, the transition to homogeneous
matter is shifted to higher densities, see also \cite{Paisthisissue}.

\begin{figure}
\resizebox{0.52\textwidth}{!}{%
\includegraphics{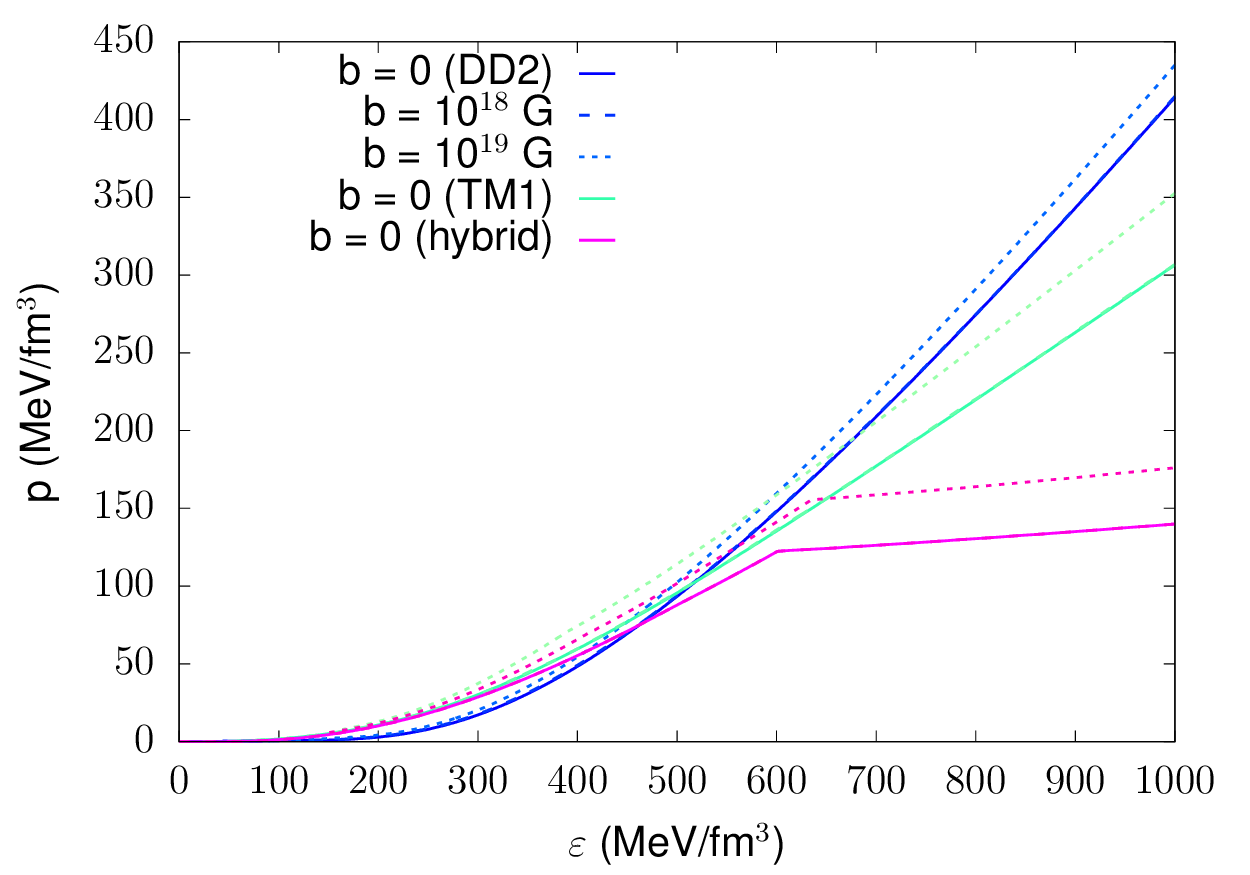}
}
\caption{The effect of magnetic field on two hadronic EoS of neutron star matter~\cite{Rabhi2008}, using the DD2~\cite{DD2} and the TM1~\cite{TM1} nuclear interaction, respectively, as well as the hybrid EoS discussed in \cite{Dexheimer2012}. 
\label{fig:eos_B_hadron}}       % Give a unique label
\end{figure}

For
homogeneous hadronic matter, be it purely nucleonic or including additional
particles such as hyperons or $\Delta$-baryons, the effect of a strong magnetic field on the EoS has been discussed for instance in
Refs.~\cite{Broderick00,Broderick02,Rabhi2008,Sinha,Haber,Dexheimer2021}
using different models for the underlying hadronic interaction. As can
be easily estimated from the value of the critical magnetic field for
the different hadrons and the energy of the interaction of the
respective magnetic moments with the magnetic field, below a field
strength of $ \sim 10^{18}$~G almost no influence on the EoS is
expected. As an example we show in Fig.~\ref{fig:eos_B_hadron} the
EoS, i.e. pressure as function of energy density for cold,
$\beta$-equilibrated matter, for two different nucleonic
models~\cite{Rabhi2008} from the family of covariant density
functional theory, DD2~\cite{DD2} and TM1~\cite{TM1}, for different
magnetic field strengths. It is obvious that indeed at $b = 10^{18}$ G
the EoS is still almost indistinguishable from the EoS without magnetic
field, whereas at $b = 10^{19}$~G the impact of the magnetic field on
the EoS becomes non-negligible.

%%%%%%%%%%%%%%%%%%%%%%%%%%%%%%%%%%%%%%%%%%%%%%%%%%%%%%%%%%%%%%%%%%%%%%%%%%%%%%
%%%%%%%%%%%%%%%%%%%%%%%%%%%%%%%%%%%%%%%%%%%%%%%%%%%%%%%%%%%%%%%%%%%%%%%%%%%%%
Since the magnetic field influences quark and hadronic matter in a
different way, it has an impact on the potential transition from
hadronic to quark matter, too. This question has been addressed by
several authors, see e.g.~\cite{Rabhi,Dexheimer2012}. In
Fig.~\ref{fig:eos_B_hadron} we show as an example the results from the
study of \cite{Dexheimer2012}. It is based on a chiral sigma model
description of both, the hadronic and the quark phase. The density for
the onset of quark matter is clearly visible from the kink in the
EoS. Actually, within the density range shown in the figure, only the
hadronic and a mixed phase are visible, whereas the pure quark phase
is present only at higher densities. In the present model, no pure
quark phase is reached in the interior of neutron stars. As can be
seen from the figure, a strong magnetic field pushes the onset of
quark matter to higher densities and pressures, thus further delaying
the appearance of a potential quark core in neutron stars. In other
studies, it is found in a similar way that a strong magnetic field
shifts the transition to quark matter to higher densities, see the
review~\cite{Menezeshere}, too.

\begin{figure}
\resizebox{0.52\textwidth}{!}{%
\includegraphics{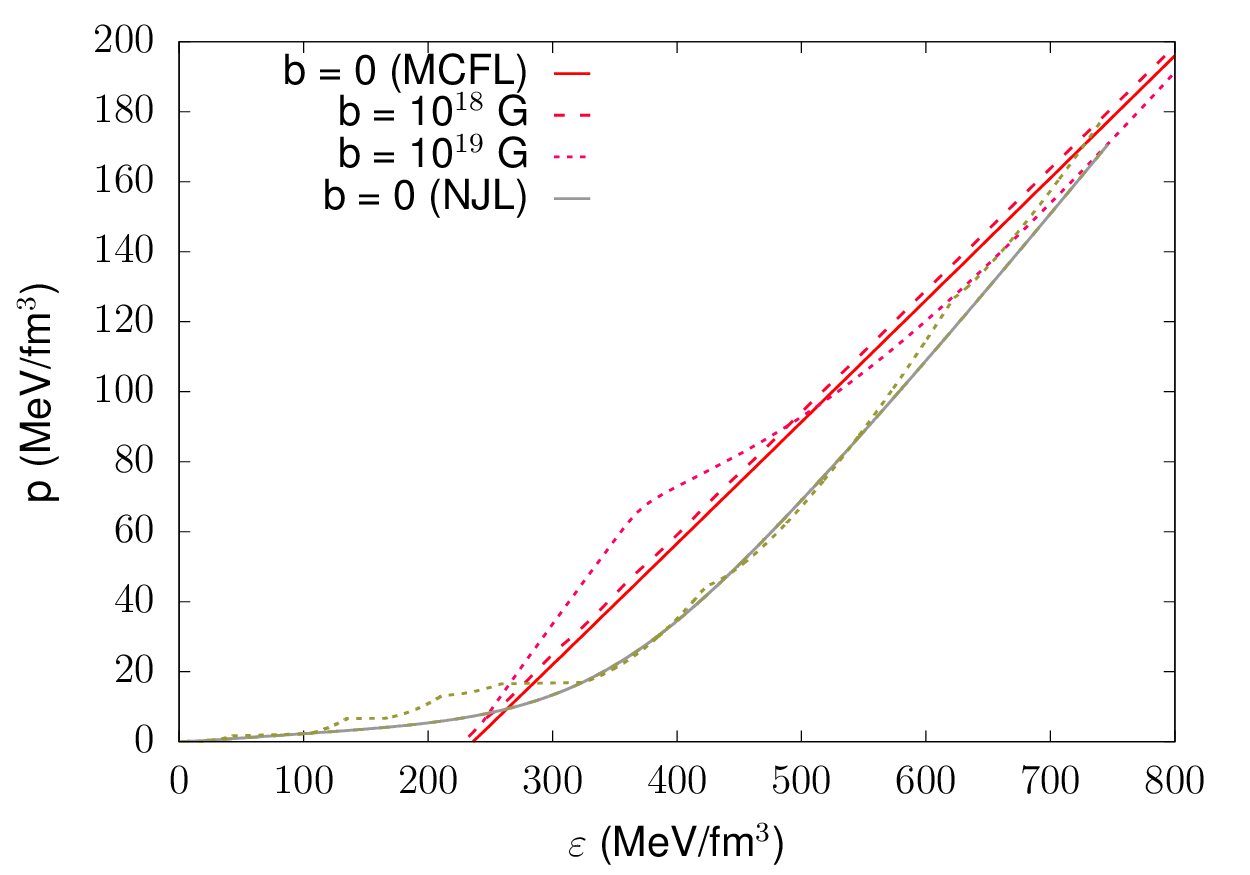}
}
\caption{The effect of magnetic field on the quark matter EoS in the NJL model discussed in \cite{Avancini} and the mCFL model discussed in \cite{Chatterjee2015}. 
\label{fig:eos_B_quark} }      % Give a unique label
\end{figure}

The influence of the magnetic field on quark matter has attracted much
interest,among others due to the impact on the complex structure of
the QCD phase diagram and phenomena such as magnetic catalysis, see
e.g. \cite{Schmitt} for a review, and several authors have considered
quark matter in compact stars with a strong magnetic field, see
e.g.~\cite{Bandyopadhyay,Noronha,Gatto,Ferrer2013,Ferrer2020} and
references therein. Two examples for influence of a strong magnetic
field on the EoS of quark matter in a compact star are shown in
Fig.~\ref{fig:eos_B_quark}. The first one~\cite{Avancini} uses the
Nambu- Jona-Lasinio (NJL) model without pairing interaction, whereas
for the second one a simple massless three-flavor MIT bag model was
applied, including a pairing interaction of NJL-type to account for
the possibility of colour superconductivity in the so-called magnetic
colour-flavor locked state (mCFL)~\cite{Chatterjee2015}, see also
\cite{Noronha}.  The most favored phase of QCD at high densities is
the colour-flavor-locked (CFL) superconducting phase and, for a
magnetic field strength of the order of the quark energy gap, such a
mCFL phase is preferred.
%%%%%%%%%%%%%%%%%%%%%%%%%%%%%%%%%%%%%%%%%%%%%%%%%%%%%%%%%%%%%%%%%%%%%%%%%%
%%%%%%%%%%%%%%%%%%%%%%%%%%%%%%%%%%%%%%%%%%%%%%%%%%%%%%%%%%%%%%%%%%%%%%%%%%%
Again, the effect of magnetic fields starts to become evident only for
very large fields above 10$^{18}$ G. For $b = 10^{19}$~G in both
models the de Haas van Alphen oscillations due to the Landau
quantisation of charged particles are clearly visible. The
dimensionless ratio $x$ of magnetisation to the magnetic field, see
Eq.~(\ref{eq:def_x}) below, is shown in Fig. ~\ref{fig:mag_mcfl} for
the mCFL model for two values of baryon number chemical
potential. Here again, the de Haas van Alphen oscillations are clearly
visible and become more and more pronounced with increasing magnetic
field strengths as expected, see also the discussion in
\cite{Noronha}.

%%%%%%%%%%%%%%%%%%%%%%%%%%%%%%%%%%%%%%%%%%%%%%%%%%%%%%%%%%%%%%%%%%%%%%%%%%%
\begin{figure}
\resizebox{0.5\textwidth}{!}{%
\includegraphics{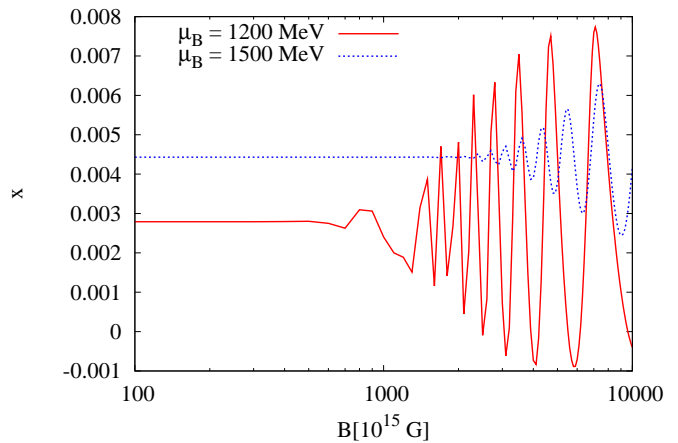}
}
\caption{Dimensionless ratio $x$, see Eq.~(\ref{eq:def_x}), of magnetisation to the magnetic field as a function of magnetic field strength $b$ in the mCFL phase showing the de Haas van Alphen oscillations \cite{Chatterjee2015}. }
\label{fig:mag_mcfl}       % Give a unique label
\end{figure}
%%%%%%%%%%%%%%%%%%%%%%%%%%%%%%%%%%%%%%%%%%%%%%%%%%%%%%%%%%%%%%%%%%%%%%%%%%%%

%% electrons in WD core?
 
%\subsubsection{Effect on crust}
%Magnetar QPOs \cite{Nandi2017}, Gabler..\\

\subsection{Effects on external structure}
\label{sec:ext}

Stellar structure is determined in General Relativity by equations describing the stationary configuration for the fluid, and the Einstein field equations. The energy- momentum tensor, containing matter properties of the star, enters the stellar structure equations as the source of the Einstein equations. 
Neglecting the coupling to the electromagnetic field, one generally assumes a perfect fluid and the energy-momentum tensor takes the following form
\begin{equation}
T_f^{\mu\nu} = (\varepsilon + p)\; u^\mu u^\nu + p\; g^{\mu\nu}~,
\label{eq:perfectfluid}
\end{equation}
where $\varepsilon$ denotes the (matter) energy density, $p$ the
pressure, and $u^\mu$ the fluid four-velocity. The EoS then relates pressure and energy density to the relevant thermodynamic quantities. In ~\cite{Chatterjee2015}, the general expression for the energy-momentum tensor in presence of an electromagnetic field was derived, starting with a microscopic Lagrangian including interaction between matter (fermions denoted by $\psi$) and the electromagnetic field $F_{\mu \nu}$
\begin{equation}
{\cal L} = - \bar{\psi}(x) (D_{\mu}\gamma^\mu + m)\psi(x) - \frac{1}{4 \mu_0} F_{\mu \nu} F^{\mu \nu}~,
\label{eq:lagr}
\end{equation}
where $D_{\mu} = \partial_{\mu} - i q A_{\mu} $ with the charge $q$ of the particle. $\mu_0$ is the vacuum permeability.
$F^{\mu \nu}$ is the field strength tensor of the electromagnetic field
\begin{equation}
F^{\mu\nu} = \partial^\mu A^\nu - \partial^\nu A^\mu ~, 
\label{e:def_fmunu}
\end{equation} 

It was demonstrated that the thermal average of the energy-momentum tensor is given by ~\cite{Chatterjee2015}
\begin{eqnarray}
\langle T^{\mu\nu}\rangle  &=& (\varepsilon + p)\; u^\mu u^\nu + p
  \; g^{\mu\nu} \nonumber \\ && + \frac{1}{2} (F^\nu_{\;\tau} M^{\tau\mu} +
F^{\mu}_{\;\tau} M^{\tau\nu} ) \nonumber \\ && 
 - \frac{1}{\mu_0} (F^{\mu\alpha} F_{\alpha}^{\,\nu} + \frac{g^{\mu\nu}}
{4} F_{\alpha\beta} F^{\alpha\beta}) ~.
\label{eq:tmunu}
\end{eqnarray}
 The first two terms on the right hand side of Eq.~(\ref{eq:tmunu}) are the perfect fluid fermionic contribution, followed by the magnetisation term and finally the electromagnetic field contributions to the energy-momentum tensor. 

According to Ohm’s Law, and assuming that the matter
has an infinite conductivity, the electric field as measured
by the fluid comoving observer must be zero, and in the fluid rest frame (FRF) only the magnetic field $b_\mu$ is nonzero. The electromagnetic field tensor can then be expressed as~\cite{Gourgoulhon}
\begin{equation}
F_{\mu\nu} = \epsilon_{\alpha\beta\mu\nu} u^{\beta} b^{\alpha}~.
\label{eq:deffmunu} 
\end{equation}
Assuming an isotropic medium and a magnetisation parallel to the magnetic field, the magnetisation tensor $M_{\mu \nu}$ can be written as 
\begin{equation}
M_{\mu \nu} =  \epsilon_{\alpha \beta \mu \nu} u^{\beta} a^{\alpha}~,
\label{eq:magtensor}
\end{equation}
where the magnetization four-vector is defined as
\begin{equation}
a_{\mu} = \frac{x}{\mu_0} b_{\mu}
\label{eq:def_x}
\end{equation}
Employing Eqs.~(\ref{eq:deffmunu},\ref{eq:magtensor}), the energy-momentum
tensor can be rewritten in the following way
\begin{eqnarray}
T^{\mu\nu}& =& T_f^{\mu\nu} 
 + \frac{1}{\mu_0} \left( - b^\mu b^\nu + (b\cdot b) u^\mu u^\nu + \frac{1}{2}
   g^{\mu\nu} (b \cdot b) \right) \nonumber \\ &&
 + \frac{x}{\mu_0}  \left( b^\mu b^\nu - (b\cdot b)( u^\mu u^\nu + 
   g^{\mu\nu})\right) ~.
\end{eqnarray}
In the absence of magnetisation, i.e. for $x=0$, this expression
reduces to the standard MHD form for the energy-momentum
tensor~\cite{Gourgoulhon}.

It is important here to point out that the magnetic field does not
induce an anisotropy in the thermodynamic matter pressure, obtained
from the derivative of the partition function. Rather magnetic fields
result in an anisotropy of the energy momentum tensor and break
spherical symmetry. Consequently with increasing strength of the
magnetic field, the shape of a magnetar departs more and more from a
spherical shape, to that of an oblate or prolate spheroid, depending
on the configuration of the field.  If the spatial elements of the FRF
energy-momentum tensor are interpreted as pressures, then there is a
difference induced by the orientation of the magnetic field, often
tabbed ``parallel'' and ``perpendicular'' pressures in the
literature. To be precise, let us consider a magnetic field pointing
in $z$-direction. The energy-momentum tensor can then be decomposed
into the sum of matter and field contributions $T_{\mu \nu} = T^{\mu
  \nu}_m + T^{\mu \nu}_B $ as follows:
\begin{eqnarray}
 T^{\mu \nu}_m = 
 \begin{pmatrix}
\varepsilon & 0 & 0 & 0\\
0 & p-ab & 0 & 0\\
0 & 0 &  p-ab & 0 \\
0 & 0 & 0 &  p
\end{pmatrix}	
\end{eqnarray}
where the first term is related to the fluid energy density and the other three non-zero terms are associated with the matter pressure in $x,y,z$ coordinates, $a$ being the magnetisation. Similarly, the electromagnetic contribution to the energy momentum tensor in the FRF becomes:
\begin{eqnarray}
 T^{\mu \nu}_B = \frac{1}{\mu_0}
 \begin{pmatrix}
\frac{1}{2}b^2 & 0 & 0 & 0\\
0 & \frac{1}{2}b^2 & 0 & 0\\
0 & 0 & \frac{1}{2}b^2 & 0 \\
0 & 0 & 0 & -\frac{1}{2}b^2~.
\end{pmatrix}	
\end{eqnarray}
The spatial elements $ T^{11} = T^{22}$ are then interpreted as
perpendicular pressure ($p_{\perp} = p - a b + b^2/2$) and the
element $T^{33}$ as parallel one ($p_{\parallel} = p- b^2/2$)~\cite{Ferrer,Paulucci,Dexheimer2014}. It should, however, be
kept in mind that these do not correspond to the thermodynamic
pressure, and have a meaning only as elements of the energy-momentum
tensor $T^{\mu \nu}$.

It should be stressed in addition, see the argumentation in \cite{Blandford} and \cite{Potekhin}, that upon computing equilibrium, the magnetisation contribution to
the energy-momentum tensor is cancelled by the Lorentz
force associated with magnetisation. Thus, although the magnetic field induces an anisotropy in the
matter part of the energy-momentum tensor, only the isotropic thermodynamic pressure $p$ is relevant for determining equilibrium. This conjecture~\cite{Blandford} has been confirmed by the derivation of the magnetostatic equilibrium equations in \cite{Chatterjee2015}, see
the discussion in Sec.~\ref{ss:equil} below.

\subsubsection{TOV equations}
For spherically symmetric neutron stars in static equilibrium, given an EoS, the global structure (e.g. mass, radius) can be calculated using the well known Tolman Oppenheimer Volkoff (TOV) equations of hydrostatic equilibrium:
\begin{eqnarray}
\frac{d \ell}{d r} &=& 4 \pi r^2 \varepsilon(r)~, \nonumber \\
\frac{d \nu}{d r} &=& \left( \ell(r) + 4 \pi r^3 p(r) \right) \frac{1}{r^2}  \left( 1-\frac{2\ell}{r} \right)^{-1}~, \nonumber \\ 
\frac{d p}{d r} &=& -\left( \varepsilon(r) + p(r) \right) \frac{d \nu}{d r}~. \label{eq:TOV}
\end{eqnarray}
Here $\ell(r)$ and $\nu(r)$ are gravitational potentials for the line element in spherical coordinates:
\begin{equation}
ds^2 = - e^{2 \nu(r)} dt^2 + \left( 1 - \frac{2\ell}{r} \right)^{-1} dr^2 + r^2 (d \theta^2 + \sin^2 \theta d \phi^2)~.
\end{equation}
The pressure and energy density profiles $p(r)$ and $\varepsilon(r)$ are related through the EoS.

The question arises whether spherical TOV equations can be applied to
describe magnetar structure, given that magnetars are highly deformed
from their spherical shape and that a magnetic field cannot be
described in spherical symmetry, as there are no magnetic monopoles.
There exist several works in the literature that attempted to compute
the mass-radius relations of strongly magnetised NSs as a first
approach using isotropic TOV equations
\cite{Strickland,Rabhi,Ferrer,Paulucci,Lopes,Dexheimer2014,Casali}. In
such works, the different components of the macroscopic
energy-momentum tensor in the FRF are used to obtain the energy
density $\varepsilon$, parallel, $p_{\parallel}$, and perpendicular,
$p_{\perp}$, pressures, respectively. However, this procedure
necessarily induces confusion, since the TOV equations only admit one
pressure.

The impossibility to apply spherically symmetric TOV equations in
presence of a magnetic fields can also be understood by inspecting the
most general solution of the equations of hydrostatic equilibrium in
general relativity for the spherically symmetric case. The coupled
system of structure equations was derived by Bowers and Liang
~\cite{Bowers}
\begin{eqnarray}
  \frac{d\ell}{d\bar r} &=& 4\pi \bar r^2 \varepsilon \nonumber\\
  \frac{d\nu}{d\bar r} &=&  \left( \ell + 4\pi \bar r^3 p_r \right)
                           \frac{1}{\bar r^2} \left( 1 - \frac{2\ell}{\bar r} \right)^{-1}
                            \nonumber\\ 
  \frac{d p_r}{d\bar r} &=& -\left(\varepsilon + p_r \right) \frac{d\nu}{d\bar r} +
                           \frac{2}{\bar r} (p_\perp - p_r) ~, 
                           \label{eq:bowers}
\end{eqnarray} 
with the most general energy-momentum tensor compatible with spherical symmetry given by 
\begin{equation}
  T^{\mu\nu} = \mathrm{diag}(\varepsilon, p_r, p_\perp,p_\perp)~,
  \label{eq:TmunuBowers}
\end{equation}
where $p_r$ and $p_{\perp}$ are the radial and tangential pressure
components. In the perfect-fluid model (\ref{eq:perfectfluid}) $p_r =
p_\perp$, and it might be tempting to cast an energy-momentum tensor
for a magnetic field pointing in $z$-direction into this form,
assuming a perfect conductor and isotropic matter. However, in the
case of the electromagnetic energy-momentum tensor $T^{\theta\theta}
\neq T^{\phi\phi}$ (see e.g. Eqs.~(23d)-(23e)
of~\cite{Chatterjee2015}), in clear contradiction with the assumption
of Bowers and Liang~(\ref{eq:TmunuBowers}) in spherical
symmetry. Further, $\lim_{r\to 0} (T^{rr} - T^{\theta\theta}) \not= 0$
and thus, the last term in Eq.~(\ref{eq:bowers}) diverges at the
origin. This discussion shows that there cannot be any correct
description of the magnetic field in spherical symmetry.

\subsubsection{Perturbative approach}
\label{sec:pert}

For low magnetic fields, Konno et al. \cite{Konno} obtained an analytical solution for the external structure using a perturbative approach. There have been also attempts \cite{Mallick} to compute the structure of neutron stars in strong magnetic fields by a simple Taylor expansion of the energy-momentum tensor and the metric around the spherically symmetric case, but these deviations are significant at ultrastrong magnetic fields relevant for magnetars. Such solutions are not applicable in the presence of large quantising magnetic fields. 
%We discuss the implications of these results in Sec.~\ref{sec:magprofile}. 
\\

\subsubsection{Numerical solutions with non-magnetised EoSs}
\label{sec:numerical}

Bocquet et al. \cite{Bocquet} were the first to include magnetic fields to rotating neutron star models, by solving coupled Einstein-Maxwell equations. Subsequently several groups performed fully relativistic numerical computations of the magnetar structure by taking into account the anisotropy of the stress-energy tensor  \cite{Cardall,Kiuchi2008,Oron,Ioka,KiuchiKotake,Yasutake,Yoshida,Bucciantini}. However, such studies assume a perfect fluid, polytrope or a realistic EoS, but do not take into account the magnetic field modifications due the quantising magnetic field, described in Sec.~\ref{sec:int}. \\

The internal composition, particularly the nuclear EoS, affects the macroscopic structure as well as observable properties of neutron stars, such as cooling and emission of gravitational waves. Correct interpretation of astrophysical data, in the era of multi-messenger astronomy, calls for the need to develop consistent numerical models of magnetars, taking into account the effect of strong magnetic fields both on the microphysics and macrophysics. Ideally, one must solve the coupled Einstein-Maxwell equations, along with a magnetic field dependent EoS, to obtain the global quantities. This is described in Sec.~\ref{sec:model}.

%%%%%%%%%%%%%%%%%%%%%%%%%%%%
\section{Numerical Models of strongly magnetised compact stars}
\label{sec:model}

In the preceding sections Sec.~\ref{sec:int} and Sec.~\ref{sec:ext}, we discussed how ultrastrong magnetic fields affect the internal and external structure of magnetars. In this section, we describe how global numerical models for the same, within the framework of general relativity, can be obtained in a consistent way.

Within the framework of general relativity, we follow \cite{BGSM} to make the assumption of a stationary, axisymmetric spacetime, in which the matter part (the energy-momentum tensor) fulfills the {\it circularity condition}. The line element in these coordinates can be written as :
\begin{eqnarray}
  {\rm d}s^2 &=& -N^2\, {\rm d}t^2 + B^2r^2\sin^2\theta \left({\rm d}\varphi -
    N^\varphi\, {\rm d}t \right)^2 \nonumber\\
  &&+ A^2\left( {\rm d}r^2 + r^2\, {\rm d}\theta^2\right),
  \label{e:def_metric}
\end{eqnarray}
where the gravitational potentials $N, N^\varphi, A$ and $B$ are functions of coordinates $(r,\theta)$. An important point of full general-relativistic models is that the magnetic field enters the sources of the gravitational field equations (Einstein equations). However, the main influence of magnetic field is in the equilibrium balance, with a contribution of a Lorentz-like force (see Sec.~\ref{ss:equil}).

%%%====================
\subsection{Maxwell equations}
\label{ss:Maxwell}

The electromagnetic field is assumed here to originate from free currents $j^\sigma$ \cite{Bocquet,Chatterjee2015}, \textit{a priori} independent from the
movements of inert mass (with 4-velocity $u^\mu$). 
This assumption is a limitation of the model, and in principle one should derive a distribution for the free currents using a multifluid approach to model the movements of free charged particles, but this is beyond the scope of this article. The four-potential $A_\mu$ that enters in the definition of the electromagnetic field tensor $F^{\mu\nu}$ through Eq.~(\ref{e:def_fmunu}), can generate simple configurations of either purely poloidal or a purely toroidal magnetic field \cite{Kiuchi2008,Frieben}. It has been suggested via analytical considerations~\cite{Tayler,Markey} and later confirmed via numerical simulations, see e.g. \cite{Braithwaite2006,Braithwaite2007}, 
that such configurations are unstable.  The instability can rapidly
rearrange the magnetic configuration of the stars into a mixed
configuration of poloidal and toroidal fields. 
Solutions in mixed
toroidal-poloidal configurations have been developed in the {\it
  Conformally Flat Condition} \cite{Bucciantini} and recently, a new
code (COCAL) has been obtained for such mixed toroidal-poloidal
configurations in the general axisymmetric (and non-circular)
spacetimes \cite{Uryu}. However in this article, we confine our
discussion to a simple configuration, a purely poloidal one. In this context it should be mentioned that the toroidal component is expected to assume higher field strengths than the poloidal one~\cite{Akgun}. However, the maximum values estimated from the observed surface fields and the virial theorem, see Sec.~\ref{sec:maxfield}, should not be exceeded in realistic situations, such that the poloidal configuration with field strengths of $10^{18}$~G and above should already give a fairly good idea about magnetic field effects on the EoS and stellar structure.

For a purely poloidal configuration, i.e., the four-potential has vanishing components $A_r=A_\theta = 0$. The electric and magnetic fields as seen by the Eulerian observer (whose four-velocity is $n^\mu$) are then defined as $E_\mu = F_{\mu\nu}\, n^\nu$ and $B_\mu =
-\frac{1}{2} \epsilon_{\mu\nu\alpha\beta}\, n^\nu\, F^{\alpha\beta}$,
with $\epsilon_{\mu\nu\alpha\beta}$ the Levi-Civita tensor associated
with the metric~(\ref{e:def_metric}). Then the non-zero components of the magnetic field are:
\begin{subequations}
  \begin{eqnarray}
    B_r & = & \frac{1}{Br^2\sin\theta}\df{A_\varphi}{\theta} \label{e:def_Br} \\
    B_\theta & = & - \frac{1}{B \sin \theta} \df{A_\varphi}{r} \label{e:def_Bt}
  \end{eqnarray}
\end{subequations}

The homogeneous Maxwell (Faraday-Gauss) equation $F_{[\mu\nu;\lambda]} = 0$
is automatically fulfilled, for the $F^{\mu\nu}$ tensor form in Eq.~(\ref{e:def_fmunu}). The inhomogeneous Maxwell (Gauss-Amp\`ere) equation 
in presence of external magnetic field is the covariant 
derivative associated with the metric (given in Eq.~\ref{e:def_metric})),
\begin{equation}
  \frac{1}{\mu_0} \nabla_\mu F^{\nu\mu} = j_{\ \mathit{free}}^\nu +
  \nabla_\mu M^{\nu\mu}~,
\end{equation}  
can then be transformed to give
\begin{equation}
  \nabla_\mu F^{\sigma\mu} = \frac{1}{1-x} (\mu_0 j_{\ \mathit{free}}^\sigma +
  F^{\sigma\mu}\nabla_\mu x)~.
\label{eq:gauss-ampere}
\end{equation}  

This equation can be expressed in terms of the two non-vanishing components of $A^\mu$, with the Maxwell-Gauss and the Maxwell-Amp\`ere equations taking the form of Poisson-like partial-derivative equations (see \cite{Chatterjee2015} for details).

\subsection{Equilibrium equations}\label{ss:equil}

The stellar equilibrium is defined by the conditions of conservation of energy and momentum. The equations can be derived from the vanishing divergence of the energy-momentum tensor:
\begin{equation}
\nabla_\mu T^{\mu\nu} = 0.
\end{equation}

From the expression of the energy-momentum tensor (Eq.~ \ref{eq:tmunu}), the equilibrium equation in the presence of electromagnetic field is:
\begin{equation}
  \nabla_\alpha T^{\alpha\beta} = \nabla_\alpha T_f^{\alpha\beta}
  -F^{\beta\nu} j^{\mathit\ free}_{\nu}  - \frac{x}{2\mu_0}
  F_{\sigma\tau} \nabla^\beta F^{\sigma\tau}~, 
\end{equation}
where $T_f^{\alpha\beta}$ represents the perfect-fluid contribution to
the energy-momentum tensor~(\ref{eq:perfectfluid}) while the second term denotes the usual Lorentz force term, arising from free currents.
\\

In the case of rigid rotation (constant $\Omega$), a first integral of the following expression is required \cite{Bocquet}
\begin{equation}
  (\varepsilon + p) \left ( \frac{1}{\varepsilon + p} \frac{\partial
      p}{\partial x^i} + \frac{1}{N}\frac{\partial N}{\partial x^i} -
    \frac{\partial \ln \Gamma}{\partial x^i}\right)  
  -  F_{i \rho} j^\rho_{\ \mathit{free}} = 0~, 
\label{eq:first-integral}
\end{equation}
where $N$ is the \textit{lapse function} from the metric defined in Eq.~(\ref{e:def_metric}) and $\Gamma$ the fluid Lorentz factor. Each term can be interpreted physically: pressure gradient, gravitational force, centrifugal force and Lorentz force.

In order to calculate this first integral, one introduces the enthalpy
per baryon and its derivatives. It can be shown that, even in the
presence of the magnetic field, the logarithm of the enthalpy per
baryon represents again a first integral of the fluid equations. 
Note that for the neutron star case with a
magnetic field in beta-equilibrium and at zero temperature, the
enthalpy is a function of both baryon density and magnetic field (with $b^2 = b_\mu b^\mu$):
\begin{equation}
 h = h(n_b, b) = \frac{\varepsilon + p}{n_b} = \mu_b~.\label{e:def_h}
\end{equation} 
Hence one gets
\begin{equation}
  \frac{\partial \ln h} {\partial x^i} = \frac{1}{h} \left(\left. \frac{\partial
        h}{\partial n_b} \right |_b \frac{\partial n_b} {\partial x^i} +
    \left. \frac{\partial
        h}{\partial b} \right |_{n_b} \frac{\partial b} {\partial x^i}\right)~.
\end{equation}
In addition, the following thermodynamic relations are valid under the present assumptions
\begin{eqnarray}
  \left.\frac{\partial h}{\partial n_b}\right |_b &=& \frac{1}{n_b}
  \frac{\partial p}{\partial n_b} \\
  \left. \frac{\partial p}{\partial b}\right|_{\mu_b} &=& a ( =
  \sqrt{a^\mu a_\mu}) = - 
  \left. \frac{\partial
      \varepsilon}{\partial b} \right|_{n_b}~.
\end{eqnarray}
And one obtains for the derivative of the logarithm of the enthalpy
\begin{eqnarray}
  \frac{\partial \ln h} {\partial x^i} &=& \frac{1}{\varepsilon + p}
  \left[\left. \frac{\partial p}{\partial n_b}\right |_b
    \frac{\partial n_b} {\partial x^i} + \left(\left. \frac{\partial
          p}{\partial b} \right |_{n_b} - a \right) \frac{\partial b} 
    {\partial x^i}\right] \nonumber \\ &=& \frac{1}{\varepsilon + p}
  \left( \frac{\partial p}{\partial x^i} - a \frac{\partial
      b}{\partial x^i} \right)~.\label{e:dlnh}
\end{eqnarray}
Assuming that matter is a perfect conductor ($A_t =
-\Omega A_\varphi$ inside the star), it is possible to
relate the components of the electric current to the electromagnetic potential $A_\varphi$, through an arbitrary function $f$, called the
{\em current function\/}:
\begin{equation}
  \label{e:current_function}
  j^\varphi - \Omega j^t = (\varepsilon + p) f\left( A_\varphi
  \right). 
\end{equation}
Under these assumptions, the Lorentz force term becomes
\begin{equation}
  \label{e:Lorentz_force}
  F_{i \rho} j^\rho_{\ \mathit{free}} = \left( j^\varphi - \Omega j^t
  \right) \df{A_\varphi}{x^i} = - \left(\varepsilon + p \right) \df{\Pi}{x^i},
\end{equation}
with
\begin{equation}
  \label{e:def_M}
  \Pi(r, \theta) = - \int_0^{A_\varphi(r,\theta)} f(x) {\rm d}x.
\end{equation}

The last term can be written in terms of the magnetic field $b^\mu$ as:
\begin{equation}
  \frac{x}{2\mu_0} F_{\mu\nu} \nabla_i F^{\mu\nu} =
  \frac{x}{\mu_0}\left( b_\mu \nabla_i b^\mu - b_\mu b^\mu u_\nu \nabla_i
    u^\nu \right) = a \df{b}{x^i},
\end{equation} 
from the expression~(\ref{eq:deffmunu}). Thus, this last term cancels with its counterpart in
Eq.~(\ref{e:dlnh}) and the first integral (\ref{eq:first-integral})
keeps exactly the same form as without magnetisation:
\begin{equation}
  \label{e:1st_integral}
  \ln h(r, \theta) + \nu(r, \theta) - \ln \Gamma(r, \theta) + \Pi(r,
  \theta) = {\rm const}.
\end{equation}

%%%==================================================
\subsection{Numerical resolution}
\label{subsec:numerics}

As discussed in Sec.~\ref{ss:Maxwell}, the methodology discussed in
this article is valid for the chosen poloidal geometry but other
configurations such as purely toroidal or mixed ones have also been
discussed elsewhere
\cite{Kiuchi2008,Frieben,KiuchiKotake,Yasutake,Bucciantini}, but
without taking into the magnetic field dependence of the EoS and
magnetisation. The Einstein-Maxwell equations can be solved within the
numerical library Lorene~\cite{Lorene}, applying spectral methods to
solve the Poisson-like partial differential equations in the 3+1
formalism. Further details about the numerical methods applied here
can be found in \textit{e.g.}  \cite{Grandclement}. The code follows
the algorithm presented by \cite{Bocquet} but the most significant
difference comes from the fact that all the required EoS variables
(e.g., $p, \varepsilon, n_b$), depend on two parameters - the enthalpy
$h$ (Eq.~\ref{e:def_h}) and the magnetic field amplitude $b =
\sqrt{b_\mu b^\mu}$. These quantities are first computed and stored in
a tabular form, which is then read by the code computing the
equilibrium global models, and a bi-dimensional interpolation using
Hermite polynomials is used, following the method described by
\cite{swesty-96}, to ensure thermodynamic consistency of the
interpolated quantities ($p(h, b), \varepsilon(h, b), n_b(h, b)$ and
$x(h, b)$). The numerical accuracy of the solutions is verified by
monitoring that the relative accuracy lies within the upper bound
defined by the {\it relativistic virial theorem}
~\cite{Bonazzola94,Gourgoulhon94}.  \\

\begin{figure}
\resizebox{0.45\textwidth}{!}{%
  \includegraphics{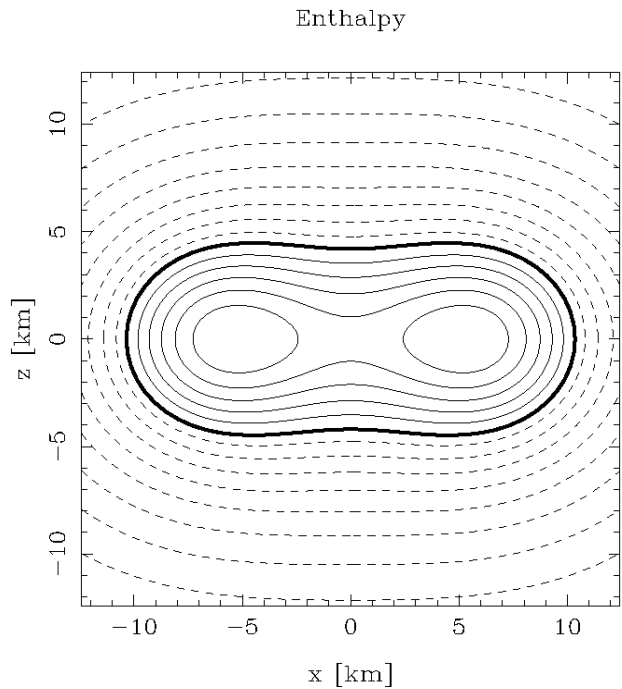}
}
\caption{The enthalpy contours in the $(x,z)$ plane for a static star of gravitational mass 2.22 $M_{sol}$ and polar field $8.16 \times 10^{17} G$  \cite{Chatterjee2015}. The bold line denotes the stellar surface.}
\label{fig:ent_maxB}       % Give a unique label
\end{figure}

\begin{figure}
\resizebox{0.45\textwidth}{!}{%
  \includegraphics{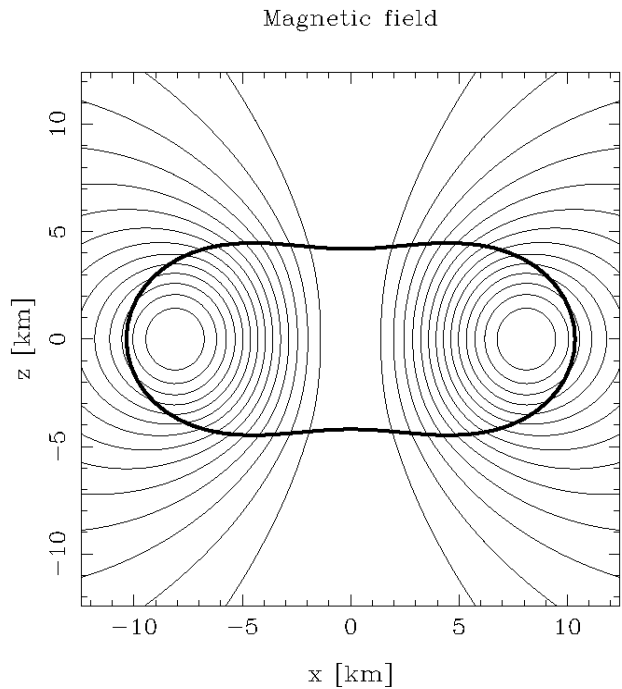}
}
\caption{The magnetic field lines for same star as in Fig.~\ref{fig:ent_maxB}. }
\label{fig:maxB}       % Give a unique label
\end{figure}

The free physical parameters entering the model are the EoS, the current function $f$ (Eq.~\ref{e:current_function}), the rotation frequency $\Omega$ and the logarithm of the central enthalpy $H_c = \log(h(r=0))$. The choice of the current function and the effect of alternative choices has been discussed in ~\cite{Bocquet}. 
%Once the equilibrium configuration has been computed, global quantities are obtained either from integration over the stellar volume (\textit{e.g.} baryonic mass $M_B$) or from the asymptotic behaviour of the gravitational field (\textit{e.g.} gravitational mass $M_G$) and of the electromagnetic field (\textit{e.g.} magnetic moment $\mathcal{M}$).
In general relativity, one must define gauge independent quantities in order to derive observables from the numerical models. The {\it gravitational mass} $M_G$ is defined as the mass perceived by a particle orbiting the star. It may be computed either as the Komar mass (using stationarity) or the ADM mass (asymptotic flatness of space) \cite{Gourgoulhon}. The {\it baryon mass} $M_B$ is the product of the baryon mass and total number of baryons in the star. The {\it circumferential equatorial/polar radius} $R_{circ}$ can be computed by integrating the line element $ds$ (Eq.~\ref{e:def_metric}) over the stellar circumference at the equator or passing through the poles. Detailed definitions and formulae can be found in Refs. \cite{Bocquet} and \cite{BGSM}.
\\
 
%%%%%%%%%%%%%%%%%%%%%%%%%%%%%%%%%%%%%%%
\section{Effect of magnetised EoS on the structure of strongly magnetised compact stars}\label{sec:mageos}

\subsection{Effect on maximum mass}

The effect of the magnetic field on the maximum mass of neutron stars
was investigated numerically in several works
e.g. \cite{Bocquet,Cardall}, assuming different EoSs (from perfect
fluid to polytropic to realistic ones). It was found that the effect
of the pure electromagnetic field is to increase the maximum mass in
neutron stars. For the maximum allowable poloidal magnetic field in
~\cite{Bocquet}, the maximum mass of neutron stars for static stars
was found to increase by 13 to 29\% (depending on the EoS) with
respect to the maximum mass of non-magnetised stars. For rotating
configurations, in most cases the magnetic field was found to be more
efficient in increasing the maximum mass than the rotation. However
such works did not take into account magnetic field modifications of
the EoS.

The investigation of the effect of magnetic field dependent EoS on the
maximum mass was done for the first time in \cite{Chatterjee2015}. The
formalism described in Sec.~\ref{sec:model} was applied to a
magnetised EoS in the mCFL (Magnetic Colour-Flavour-Locked) phase, and
the coupled Einstein-Maxwell equations were solved numerically
\cite{BGSM,Bocquet}. The numerical solution was obtained within the
numerical library Lorene \cite{Lorene} using spectral methods
\cite{Grandclement} as described in Sec.~\ref{subsec:numerics}. The
construction of numerical models for isolated rotating neutron stars
using this formalism was also discussed in \cite{Novak2015}.

It was demonstrated in accordance with earlier results
~\cite{Bocquet,Cardall}, that on increasing the magnetic field, the
Lorentz force exerted by the field deviates the stellar shape more and
more from spherical symmetry (see e.g. Fig.~\ref{fig:ent_maxB} and
Fig.~\ref{fig:maxB}). Fig.~\ref{fig:ent_maxB} also demonstrates the
limitation of the code, as the numerical scheme breaks down when the
stellar shape is too strongly deformed from spherical to a torus-like
form \cite{Cardall}. This limit is reached when the magnetic pressure
at the star's centre becomes comparable to the fluid pressure and has
been numerically explored by~\cite{Cardall}.  

\begin{figure}[h!]
\resizebox{0.52\textwidth}{!}{%
  \includegraphics{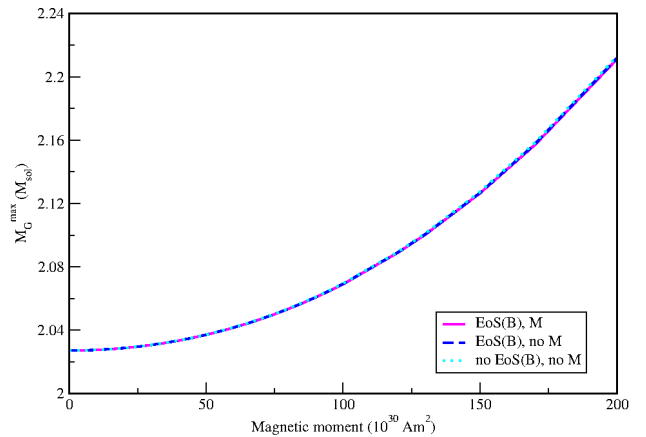}
}
\caption{NS maximum mass for models with and without magnetic field dependent EoS and magnetisation}
\label{fig:mmmgmax}      
\end{figure}

\begin{figure}[h!]
\resizebox{0.52\textwidth}{!}{%
  \includegraphics{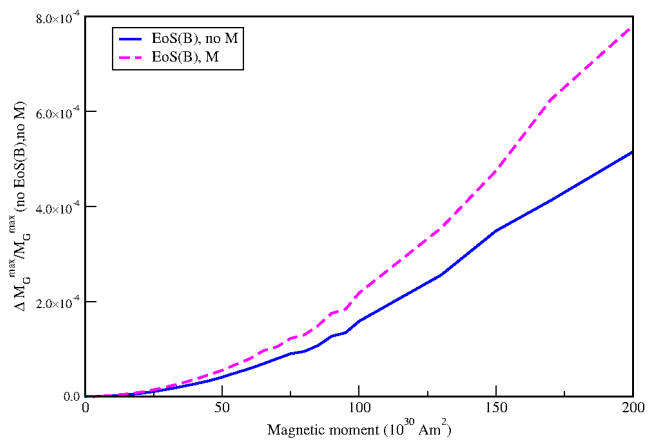}
}
\caption{Relative difference in stellar mass for the models in Fig.~\ref{fig:mmmgmax}}
\label{fig:mgratio}       
\end{figure}

In Fig.~\ref{fig:mmmgmax} and Fig.~\ref{fig:mgratio}, NS masses are
plotted as a function of the magnetic moment, which is a constant of
motion and varies linearly as the stellar magnetic field (see
\cite{Chatterjee2015,Bocquet} for discussions).  On comparison of the
models with and without magnetic field dependent EoS and
magnetisation, it was demonstrated in \cite{Chatterjee2015} that the
relative difference between the maximum masses is negligible if
magnetic fields are taken into account consistently in the modelling of
magnetars. The same conclusion was also drawn for the variation of the
stellar compactness $$C = \frac{M_G}{R_{circ}}$$ as a function
of magnetic moment, where $R_{circ}$ is the circumferential equatorial
radius \cite{BGSM}. The effect of anomalous magnetic moment was not
included in this work, as its influence on the EoS and global
structure is estimated to be small \cite{Strickland,Rabhi}. Additional
magnetic field effects such as the ferromagnetic instability
\cite{Maruyama,Vidaurre} or magnetic catalysis \cite{Haber}, if
present in neutron stars, could enhance the effect of magnetic field
on the EoS and therefore the maximum mass, too.

%%%%%%%%%%%%%%%%%%%%%%%%%%%%%%%%%
\subsection{Effect on Radius and Crust-Core properties}

Magnetars display a large number of peculiar phenomena, such as
anti-glitches, bursts and oscillations that challenge our theoretical
understanding of these objects. In Sec.~\ref{sec:int} we discussed
that in the presence of quantising magnetic fields, the motion of
charged particles are affected, which in turn, alters the EoS as well
as the properties of the crust-core boundary. Moreover, recent studies
\cite{Pons2013,Piekarewicz2014} indicate that the crust-core
transition region and the thickness of the crust play an important
role during the cooling process of a magnetar and the emission of
gravitational waves.

%\begin{figure}
%\resizebox{0.45\textwidth}{!}{%
%  \includegraphics{mrtovzdunik_sly4_allb.eps}
%  \resizebox{0.52\textwidth}{!}{\rotatebox{270}%
%{\includegraphics{mrtovzdunik_sly4_allb.eps}}
%}
%\caption{Mass-radius relation using the isotropic TOV approximation for different field strengths for SLy4 EoS }
%\label{fig:mrtov_sly4}       
%\end{figure}

As discussed in the previous section the main influence on the maximum
mass of a neutron star in a strong magnetic field does arise from the
magnetic pressure itself, the modification of the EoS and
magnetisation contributing only marginally for field strength below
$10^{19}$ G. From the discussion in Sec.~\ref{sec:int}, we expect only
a moderate influence of the magnetic field effects on the EoS on the
neutron star radii, too.  However, it has been discussed
\cite{Fang2016,Fang2017} that the magnetic field has a non-negligible
effect on the dynamical spinodals and hence the crust-core phase
transition, which in turn should affect the crust-thickness.  This
question has been addressed in \cite{Fang2016,Fang2017}, however,
applying the spherically symmetric TOV approximation me\-thod to
compute the star's structure, which is not applicable in the presence
of a magnetic field.

\begin{figure}[h!]
\resizebox{0.45\textwidth}{!}{\rotatebox{270}%
{\includegraphics{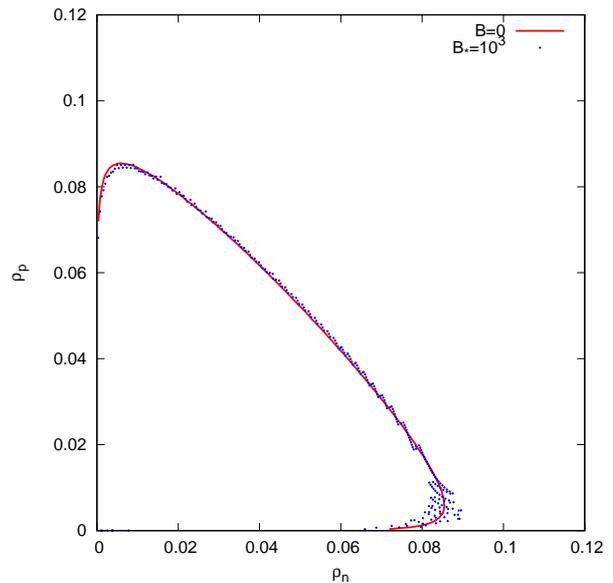}}
}
\caption{Thermodynamic spinodals for relative magnetic field strength $B_* = B / B_e^c = 10^3$, with $B_e^c = 4.4 \times 10^{13}$~G being the quantising electron field. The continuous red line gives the spinodal envelope in the absence of magnetic field~\cite{ChatterjeeJCAP2019}.}
\label{fig:tdspinbst1e3}       
\end{figure}

\begin{figure}[h!]
\resizebox{0.45\textwidth}{!}{\rotatebox{270}%
{\includegraphics{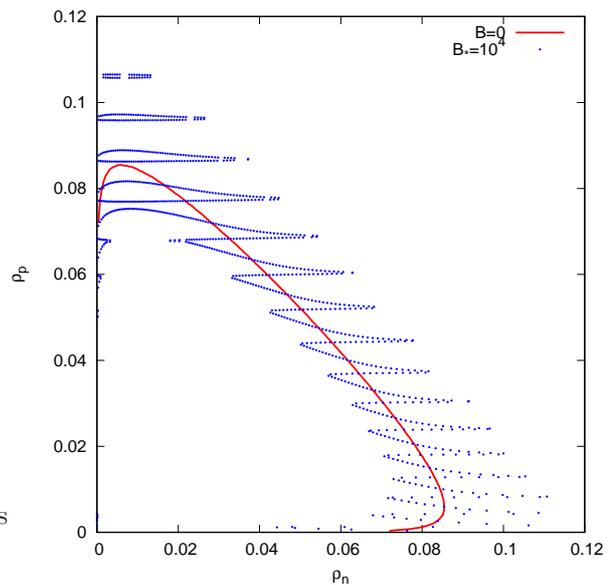}}
}
\caption{Same as Fig.~\ref{fig:tdspinbst1e3} for relative magnetic field strength $B_* = B / B_e^c = 10^4$, with $B_e^c = 4.4 \times 10^{13}$~G being the quantising electron field ~\cite{ChatterjeeJCAP2019}. }
\label{fig:tdspinbst1e4}       
\end{figure}

In \cite{ChatterjeeJCAP2019}, the problem was revisited within the
framework of density functional theory, using a ``Meta-Modelling"
technique directly related to the empirical parameters that can be
constrained through nuclear physics experiments \cite{MM}. The
numerical formalism described in Sec.~\ref{sec:model} was applied to
determine the influence of strong magnetic fields on magnetar radii
and crust thickness in non-spherical magnetars. First a study of the
influence of strong magnetic fields on thermodynamical and dynamical
spinodals was performed. The effect of different magnetic field
strengths on the thermodynamical spinodals is shown in
Figs.~\ref{fig:tdspinbst1e3} and ~\ref{fig:tdspinbst1e4}. It was
concluded that the magnetic field severely modifies the structure of
the phase transition region, leading to a non-negligible modifications
in the density and pressure of the transition from that of the zero
magnetic field case, confirming previous results.  

The effect of magnetic fields on the crust-core phase transition was
then studied with a chosen reference Skyrme SLy4 EoS
\cite{DouchinHaensel}, a commonly used EoS in astrophysics.  The ideal
way to calculate the structure of strongly magnetised neutron stars is
to solve the Einstein-Maxwell equations self-consistently with
magnetic field dependent EoS, as described in
Sec.~\ref{sec:model}. Employing this method, mass-radius relations of
magnetised neutron stars were calculated for full numerical
solutions. The unified EoSs for describing both the crust and the core
were constructed within the same ``Meta-Model" scheme
\cite{MM,Chatterjee2017,Carreau,Antic}. The magnetic field dependence
is only included within the core EoS, assuming a non-magnetised
crust. It was demonstrated that the results for previously applied TOV approximation and the fully consistent numerical
solutions vary widely due to the different techniques adapted. For low
magnetic fields the mass-radius relation resembles closely the
zero-field case. With increasing magnetic fields, it departs strongly
from the TOV solution.
%(see Fig.~\ref{fig:mrtov_sly4} vs Fig.~\ref{fig:mrcompmm_sly4})
As was already pointed out in \cite{Chatterjee2015,Chatterjee2017},
this is mainly due to the pure magnetic field contribution and not
that of the effect of the magnetised EoS. The other main difference is
that for the isotropic TOV calculations, the pure field contribution
is a constant, whereas that of the full numerical computation is a
profile, generated via a current function. This will be elaborated
further in Sec.~\ref{sec:magprofile}. The model dependence of the
results was also tested with two other reference nuclear interaction
models, TM1 and Bsk17, and the results obtained were qualitatively
similar. It was therefore concluded from this study that a full
numerical formalism is inevitable for the calculation of radii and
crust thickness strongly magnetised neutron stars.

\begin{figure}
\resizebox{0.55\textwidth}{!}{%
  \includegraphics{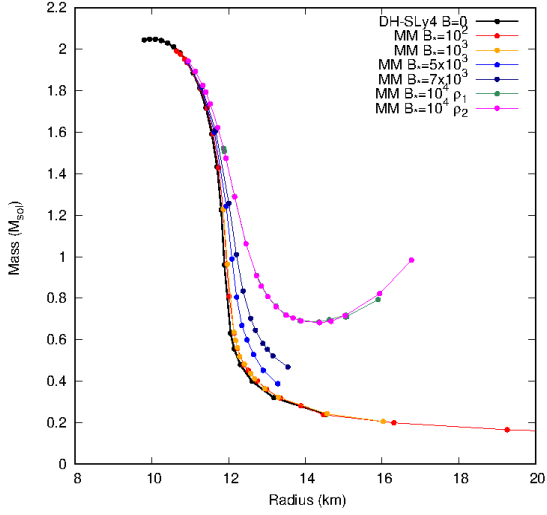}
}
\caption{Gravitational mass as a function of circumferential equatorial radius for the SLy4 EoS using full numerical calculation. $B_* = B / B_e^c$, with $B_e^c = 4.4 \times 10^{13}$~G being the quantising electron field.}
\label{fig:mrcompmm_sly4}       
\end{figure}

%%%%%%%%%%%%%%%%%%%%%%%%%%%%
\section{Magnetic field distribution}
\label{sec:magprofile}

In Sec.~\ref{sec:int}, we already underlined how quantising magnetic fields affect particle populations and their motions. In order to understand how this may affect the nuclear EoS and observable properties such as transport properties, one needs to know the maximum amplitude of the magnetic field at a given depth in the neutron star.  \\

The initial attempts involved assuming an arbitrary magnetic field profile connecting the magnetic field norm $b$ with the surface field $b_s$ and the central field $b_c$ \cite{Bandyopadhyay}:
\begin{equation}
b (n_B/n_0) = b_s + b_c [1 - \exp( - \beta (n_B/n_0)^{\gamma})]~,
\label{eq:bprofile1}
\end{equation}
with parameters $\beta$ and $\gamma$, chosen to obtain the required values of the maximum field at the centre and at the surface. This profile was subsequently used in more than 100 publications (e.g. \cite{Rabhi,Dexheimer2012,Menezes2009a,Menezes2009b}). An improvement over this formulation was attempted by others, trying to motivate a better choice of parameters \cite{Dexheimer2012,Lopes2015,Dexheimer2017a,Dexheimer2017b}.
Lopes and Menezes \cite{Lopes2015} introduced a variable magnetic field, depending on the energy density rather than on the baryon number density:
\begin{equation}
b = b_c \left( \frac{\epsilon_M}{\epsilon_0} \right)^{\gamma} + b_s~,
\label{eq:bprofile2}
\end{equation}
where $\epsilon_M$ is the energy-density of the matter contribution,
$\epsilon_0$ is the central energy density of the maximum mass
non-magnetic neutron star and $\gamma$ is a positive chosen parameter, arguing that
this formalism reduces the number of free parameters from two to
one. This profile was then applied in the calculation of the anisotropic shear stress tensor (see \cite{Bednarek,Menezes2016}), with the pressure components ``averaged" in the form diag$(b^2/24 \pi,b^2/24 \pi,b^2/24 \pi)$ but within a spherically symmetric TOV system. There have also been suggestions of the magnetic field profile being a function of the baryon chemical potential \cite{Dexheimer2012} as:
\begin{equation}
b(\mu_B) = b_s + b_c \left[ 1-\exp(\beta \frac{(\mu_B-938)^\alpha}{938} ) \right]~,
\end{equation}
with $\alpha=2.5$, $\beta=-4.08 \times 10^{-4}$ and $\mu_B$ given in
MeV. In contrast to the profiles in
Eqs.~(\ref{eq:bprofile1},\ref{eq:bprofile2}), such a formula avoids
that a phase transition induces a discontinuity in the effective
magnetic field.  Dexheimer et al. \cite{Dexheimer2017a,Dexheimer2017b}
also performed a fit to the shapes of the magnetic field profiles in
the polar direction as a function of the chemical potentials (as in
\cite{Dexheimer2012}) by quadratic polynomials instead of exponential
ones as
\begin{equation}
b(\mu_B) = \frac{(\alpha+\beta\mu_B + \gamma\mu_B^2)}{b_c^2} \mu ~,
\end{equation}
where $\alpha,\beta,\gamma$ are coefficients determined from the
numerical fit.  In Ref.~\cite{Mallick}, a density dependent profile is
applied within a perturbative axisymmetric approach similar to the
Hartle and Thorne formalism for slowly rotating stars. It remains,
however, that the stellar deformation due to the magnetic field
implies that such a density (or equivalent) dependent profile should depend
on the direction, thus will appear different looking in the polar or
the equatorial direction.  

It was later demonstrated in \cite{MenezesAlloy} that such {\it ad
  hoc} field distributions are unphysical, as they do not satisfy
Maxwell's equations. This can also be understood from the fact that
assuming such profiles in a spherically symmetric star would
incorrectly imply a purely monopolar magnetic field distribution. The
ideal way of achieving this would be to solve the coupled
Einstein-Maxwell equations described in Sec.~\ref{sec:model}, provided
with a magnetic field dependent EoS. In \cite{Chatterjee2019}, a
numerical scheme was developed to achieve the above. In order to
estimate the value of the magnetic field strength at a given stellar
depth to test its potential effect on matter properties, this work
provided a ``universal'' magnetic field strength profile from the
surface to the interior obtained from the field distribution in a
fully consistent numerical calculation.  

\begin{figure}
\resizebox{0.5\textwidth}{!}{\rotatebox{270}%
{\includegraphics{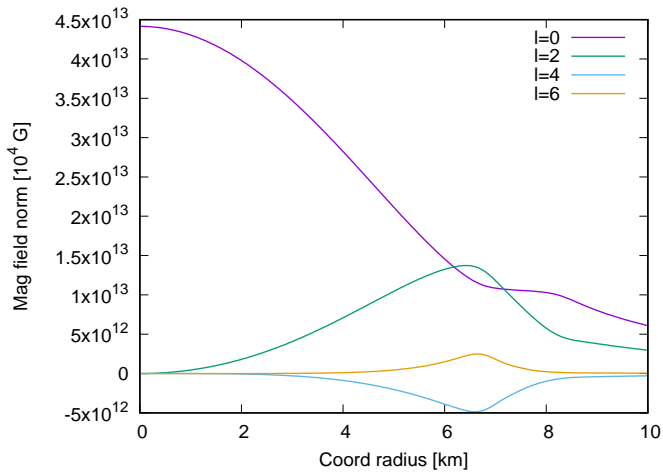}}
  }
\caption{First four even multipoles of the magnetic field norm as functions of the coordinate radius $r$ \cite{Chatterjee2019}. }
\label{fig:multi_l}       % Give a unique label
\end{figure}

\begin{figure}
\resizebox{0.5\textwidth}{!}{\rotatebox{270}%
{\includegraphics{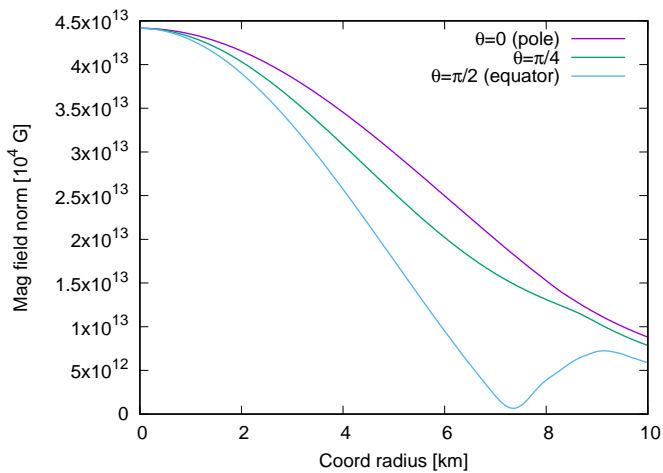}}
  }
\caption{Magnetic field norm as functions of the coordinate radius $r$ for different angular directions ($\theta=0,\pi/4,\pi/2$) \cite{Chatterjee2019}. }
\label{fig:coupes}       % Give a unique label
\end{figure}

\begin{figure}
\resizebox{0.52\textwidth}{!}{%
  \includegraphics{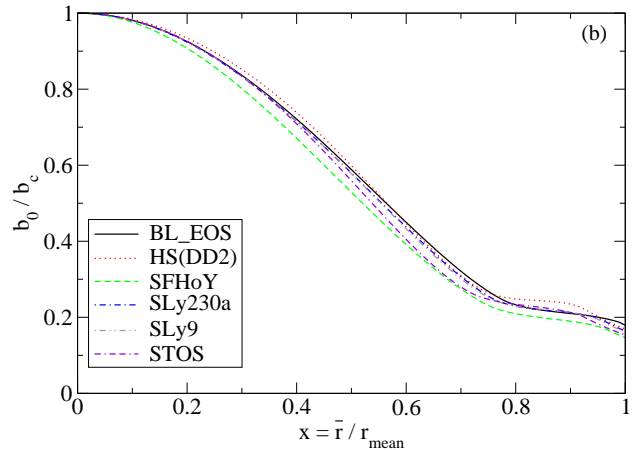}
}
\caption{ Generic profile for monopolar part of the magnetic field for different EoSs \cite{Chatterjee2019}. }
\label{fig:uprofeos}       % Give a unique label
\end{figure}

Using the full numerical solution described in Sec.~\ref{sec:model}, an extensive investigation of the magnetic field configurations was performed in \cite{Chatterjee2019}, for varying mass, magnetic field and EoS. Instead of the magnetic field (which transforms as a vector), a multipolar decomposition of the magnetic field norm (which transforms as a scalar),
\begin{equation}
 b (r, \theta) \simeq \sum_{l=0}^{l_{max}} b_l(r) \times Y_l^0(\theta)
 \label{eq:norm}
\end{equation}
was obtained, where $Y_l^m(\theta,\phi)$ are spherical harmonic functions. The first coefficients $b_l$ are displayed in Fig.~\ref{fig:multi_l}. 
It was shown (see Fig.~\ref{fig:coupes}) that the monopolar (spherically symmetric) term $b_0(r)$ is dominant. 
The profile $b_0(r)$ can be fitted by a polynomial as a function of the stellar radius:
\begin{equation}
b_0(x) = b_c \times (1-1.6 x^2 - x^4 + 4.2 x^6 - 2.4 x^8)~,
\label{eq:fit}
\end{equation}
where $x=\bar{r}/r_{mean}$ is the ratio of the stellar radius in Schwarzschild coordinates and the mean (or areal) radius. When a neutron star gets distorted from spherical symmetry, it is not possible to define a unique radius, and therefore the description in terms of the Schwarzschild coordinate radius $\bar{r}$ is a convenient one. As evident from Fig.~\ref{fig:uprofeos}, this generic eighth-order polynomial fit applies well for a wide variety of hadronic EoSs (HS(DD2), SFHoY, STOS, BL\_EOS, SLy9, SLy230a). It was thus demonstrated that the generic profile for the monopolar component of the magnetic field norm is fairly independent of a wide choice of mass, central magnetic field and hadronic EoSs (with the exception of polytropic and quark matters EoSs), and therefore can be taken as a ``universal profile''. 
\\

This magnetic field profile, derived from realistic computations, can then be applied by nuclear physicists to estimate magnetic field strength inside the neutron star, for studies of its influence on the nuclear composition. As has been shown by Chatterjee et al.~\cite{Chatterjee2019}, the star's radius is best determined by solving the simple TOV system~(\ref{eq:TOV}), thus getting the information needed to use this universal profile.

%%================================

\section{Summary and future prospects}
\label{sec:conc}
In this review we discussed recent developments in techniques to
calculate the global structure of strongly magnetised neutron stars
consistently within the general relativistic framework. We explained
why isotropic hydrostatic equilibrium equations, used to describe non-
magnetised stars, are no longer applicable in presence of an
ultrastrong magnetic field that significantly deforms the star. To
obtain the global structure, we therefore propose numerically solving
equations of stellar equilibrium obtained from conservation of
the energy-momentum tensor coupled to Einstein-Maxwell equations
\cite{Chatterjee2015}. Several of the numerical tools (e.g. \textsc{Lorene},
\textsc{XNS}) are publicly available and have been successfully applied in
subsequent works to study global properties of strongly magnetised
neutron stars as well as white dwarfs using several realistic
equations of state \cite{Chatterjee2017b,Franzon,Otoniel}.

The drawback of the numerical method elaborated in this article and
implemented in \textsc{Lorene} is in the choice of the configuration
of the magnetic field geometry, as only a purely poloidal
configuration is considered. In general also toroidal component in the
magnetic field should appear, even with a higher amplitude than the
poloidal one, see e.g. \cite{Akgun}. The effect of purely toroidal fields on the structure of
neutron stars was investigated
\cite{Kiuchi2008,Frieben,KiuchiKotake,Yasutake,Yoshida} for polytropic
EoSs. However the maximum central magnetic field being the same (as
estimated from the virial theorem) the effect on the maximum neutron
star mass will not be significantly different qualitatively for both
configurations. Solutions in mixed toroidal-poloidal configurations
have been developed in the {\it Conformally Flat Condition}
\cite{Bucciantini}, implemented in the \textsc{XNS} code, and recently,
a new code (COCAL) has been obtained for such mixed toroidal-poloidal
configurations in the general axisymmetric (and non-circular)
spacetimes \cite{Uryu}.

It was demonstrated as an application of the consis\-tent
formalism, how the maximum mass of a neutron star is negligibly
modified due to the magnetic field dependence of the microscopic EoS,
even for the highest observed neutron star magnetic fields. We also
discussed the results of the investigation of the influence of strong
magnetic fields on the crust-core phase transition properties, and
consequently on the radius and crust thickness of a neutron star
\cite{ChatterjeeJCAP2019}. It was concluded that large magnetic fields
strongly modify the crust-core phase transition. Moreover, a comparison
with the results of isotropic TOV models convincingly showed that full
numerical solutions are inevitable for the structure of strongly
magnetised neutron stars.

We further discussed an application of the full numerical solution in
constructing a ``universal" magnetic field distribution with radial
depth in neutron stars \cite{Chatterjee2019}, in place of arbitrary
magnetic fields commonly adopted in the literature, to estimate the
influence of strong magnetic fields on the NS interior
composition. The proposed fit parametrisation, obtained from the full
numerical structure calculations, may serve as a useful tool for
nuclear physicists.

%%%%%%%%%%%%%%%%%%%%%%%%%%%%%%%%%%%%%%%%%%%%%%
\section{Acknowledgements}
D.C. would like to thank Veronica Dexheimer 
and Constan\c{c}a Provid\^encia
for kindly providing the data for Figs.~\ref{fig:eos_B_hadron} and \ref{fig:eos_B_quark} 
%and Suprovo Ghosh for some of the figures 
on equations of state in this paper.

%%%%%========================= GUIDELINES 
\iffalse

\begin{figure*}
\vspace*{5cm}       % Give the correct figure height in cm
\caption{Please write your figure caption here}
\label{fig:2}       % Give a unique label
\end{figure*}
\begin{table}
\caption{Please write your table caption here}
\label{tab:1}       % Give a unique label
\begin{tabular}{lll}
\hline\noalign{\smallskip}
first & second & third  \\
\noalign{\smallskip}\hline\noalign{\smallskip}
number & number & number \\
number & number & number \\
\noalign{\smallskip}\hline
\end{tabular}
% Or use
\vspace*{5cm}  % with the correct table height
\end{table}
%
\fi
%%%%%%%%%%%%%%%%%%%%%%%%%%%%%%%%%%%%%%%%%%%%%%%%%%%
% BibTeX users please use
% \bibliographystyle{}
% \bibliography{}

\end{document}